\documentclass[twocolumn]
{aastex7}
\renewcommand{\thefigure}{\arabic{figure}}

\shorttitle{Exoplanet Host Metallicities with TRES}
\shortauthors{Rodr\'iguez Mart\'inez et al.}

\begin{document}

\title{A Uniform Determination of the Bulk Metallicities and Alpha Enrichments of Confirmed Exoplanet Systems with TRES}

\author[orcid=0000-0003-1445-9923]{Romy Rodr\'iguez Mart\'inez}
\affiliation{Center for Astrophysics \textbar \ Harvard \& Smithsonian, 60 Garden Steet, Cambridge, MA 02138, USA}
\email[show]{romy.rodriguez@cfa.harvard.edu}  

\author[orcid=0000-0002-1533-9029]{Emily K. Pass}
\affiliation{Kavli Institute for Astrophysics and Space Research, Massachusetts Institute of Technology, Cambridge, MA 02139, USA}
\email{epass@mit.edu} 

\author[orcid=0000-0002-1617-8917]{Phillip A. Cargile}
\affiliation{Center for Astrophysics \textbar \ Harvard \& Smithsonian, 60 Garden Steet, Cambridge, MA 02138, USA}
\email{pcargile@cfa.harvard.edu} 

\author[orcid=0000-0003-0741-7661]{Victoria DiTomasso} 
\affiliation{Center for Astrophysics \textbar \ Harvard \& Smithsonian, 60 Garden Steet, Cambridge, MA 02138, USA}
\email{victoria.ditomasso@cfa.harvard.edu}

\author[orcid=0000-0002-9003-484X]{David Charbonneau}
\affiliation{Center for Astrophysics \textbar \ Harvard \& Smithsonian, 60 Garden Steet, Cambridge, MA 02138, USA}
\email{dcharbonneau@cfa.harvard.edu}

\author[orcid=0000-0003-3773-5142]{Jason D. Eastman}
\affiliation{Center for Astrophysics \textbar \ Harvard \& Smithsonian, 60 Garden Steet, Cambridge, MA 02138, USA}
\email{jason.eastman@cfa.harvard.edu}

\author[0000-0001-9911-7388]{David W. Latham}
\affiliation{Center for Astrophysics \textbar \ Harvard \& Smithsonian, 60 Garden Steet, Cambridge, MA 02138, USA}
\email{dlatham@cfa.harvard.edu}

\begin{abstract}

We present a uniform spectroscopic characterization of 625 F, G, and K stars hosting 859 confirmed exoplanets using high-resolution archival optical spectra from the Tillinghast Reflector Echelle Spectrograph (TRES). We use the neural network spectral code {\tt uberMS}, which combines spectra with broadband photometry to estimate precise and accurate stellar parameters. We determine stellar effective temperatures, surface gravities, radii, luminosities, projected rotational velocities, [Fe/H] abundances, and [$\alpha$/Fe] enrichments for most confirmed planet hosts observed by TRES. This uniform catalog can be used for a broad range of astrophysical studies, particularly to explore links between stellar [$\alpha$/Fe] and a suite of observed exoplanet properties. Combining our metallicity measurements with galactic kinematics, we identify 58 planet hosts that are likely members of the thick disk. We investigate the chemical environments of giant-planet formation by comparing the [$\alpha$/Fe] distributions of giant-planet host stars across different metallicity regimes. We find that subsolar metallicity giant-planet hosts are significantly enhanced in [$\alpha$/Fe] relative to Fe-rich giant-planet hosts and to the average Fe-poor field star, at high statistical significance. This suggests that enhanced $\alpha$-element abundances may partially compensate for low-Fe content and thus enable the formation of giant planets in metal-poor environments. We additionally compare the [$\alpha$/Fe] distributions of single- and multi-planet hosts and find modest evidence that $\alpha$-enhanced stars may preferentially host multi-planet systems. Finally, we recover previously observed trends between stellar metallicity and planetary eccentricity. 

\end{abstract}


\section{Introduction} 

Our understanding of the physical properties, formation and evolution of exoplanets critically depends upon our knowledge of the properties of their host stars. For example, a planet's bulk properties, such as its mass and radius, can only be as precisely measured as the mass and radius of its parent star. This fact, and the rapidly growing number of confirmed exoplanets in recent years has therefore motivated significant efforts to better characterize stars. The advent of space missions like 
{\it Gaia} \citep[][]{Gaia:2018, GaiaDR3}, which have provided high-precision parallaxes for millions of stars in the Milky Way, and high-resolution spectroscopic surveys like the Apache Point Observatory
Galactic Evolution Experiment (APOGEE; \citealt{Majewski:2017}) and {\it Gaia}-ESO \citep{Smiljanic:2014}, to name a few, have enabled us to measure the fundamental properties of large numbers of stars with unprecedented precision.

One property of particular interest is stellar metallicity: the abundance of elements heavier than H and He in a star's photosphere,  as it is considered a proxy for the solid surface density in the primordial planet-forming protoplanetary disk (when combined with the stellar mass; \citealt{Dawson:2015}). Bulk metallicity strongly correlates with planet occurrence rate \citep[e.g.][]{Fischer:2005, Buchhave:2012, Wang:2015, Petigura:2018}, planet composition \citep{Schulze:2021, Adibekyan:2021}, orbital properties \citep{Dawson:2013}, and possibly system architecture as a whole (e.g., \citealt{BrewerMultis:2018, Lozovsky:2025}). Notably, it has been robustly established that giant planets occur more frequently around metal-rich stars, and that metal-rich stars host a larger diversity of exoplanets, although the link between metallicity and the frequency of small planets is less clear \citep{Petigura:2018}. In addition, evidence for connections between planet multiplicity and metallicity have recently begun to emerge, although with some debate in the literature \citep{Weiss:2018, Munoz-Romero:2018, Anderson:2021, RodriguezMartinez:2023b}. 

Unfortunately, only a fraction of exoplanet hosts have reported abundances of elements beyond iron \citep{Akeson:2013}, although there has been a growing body of work in this area, \cite[e.g.,][]{Hinkel:2014, Petigura:2017,Brewer:2016, BrewerFischer2018, Kolecki:2022, Polanski:2022, Gore:2024, Du:2024, Sharma:2024, Sharma:2025}. Furthermore, modern studies of elemental abundances employ a variety of different techniques, with results often depending on the choices made on atomic data, model atmospheres, or radiative transfer codes. As a result, published stellar abundances are highly heterogeneous, which poses significant challenges to uncovering more subtle, fundamental links between exoplanets and the detailed chemical composition of their host stars. Drawing robust, population-level conclusions about planetary systems fundamentally requires large samples of homogeneously characterized planet hosts.

One of the applications of such a large, uniform sample is the exploration of planet demographics across galactic stellar populations. The Milky Way can be approximately split into three main stellar components: the thin disk, the thick disk, and the halo \citep{Gilmore:1983, Chiba:2000}. The vast majority of stars (and known planet hosts) are in the thin disk. Stars in the thick disk are characterized by having higher scale heights and space velocity dispersions, and they are considerably older than thin disk stars (e.g, \citealt{Bensby:2003, Bensby:2005, Xiang:2022}). Most notably, thick-disk stars have significantly distinct metallicity distributions, with higher [$\alpha$/Fe] and lower [Fe/H]. These stars are thought to be enriched in $\alpha$-elements produced from the explosions of Type II supernovae, reflecting their formation at early times before significant iron enrichment from Type Ia supernovae. The origin of the thick disk and its properties is still debated, but some theories include mergers with minor satellites, which result in either the accretion of stars from those satellites or star formation triggered by such mergers (e.g., \citealt{Helmi:2018}). 

Because most photometric and spectroscopic planet searches have typically focused on the nearest stars, there are currently not as many confirmed planets in the thick disk compared to the thin disk. Yet, their differences in age and metallicity could result in significantly different types of planetary systems around them and could thus provide powerful clues about planet formation. Indeed, there is emerging evidence that location in the Galaxy may at least partially affect planet frequency and composition. For example, \cite{Dai:2021} and \cite{Bashi:2022} found lower occurrence rates of small planets at higher relative galactic velocities of the hosts, and at higher galactic amplitudes, respectively. In addition, \cite{Zink:2023} recently found additional evidence for reduced occurrence rates of small planets at high galactic amplitudes, while \cite{Sagear:2025} found that singly-transiting kinematic thick-disk planets exhibit higher eccentricities than those in the thin disk. This deficit in planet occurrence rates in the thick disk may be attributed to a combination of low metallicity (fewer initial rock-forming solids) and high UV radiation from nearby stars resulting in disk photoevaporation \citep{Hallatt:2025}.

Planet composition itself may also be deeply linked to galactic chemical evolution. In particular, the metal-poor, $\alpha$-enhanced environments in the thick disk could lead to the formation of rocky planets with smaller Fe cores and larger magnesium silicate mantles (i.e., smaller core mass fractions; \citealp{Santos:2017, Behmard:2025, Ferreira:2025}). However, the number of confirmed small planets in the thick disk remains too small to robustly determine the compositional properties of this population, and additional discoveries in this region will be essential for understanding planet formation and evolution in a broader galactic context.  

In this paper, we provide a uniform catalog of precise and accurate stellar parameters --including iron and $\alpha$-process enrichment-- for over 600 stars with confirmed exoplanets that have been observed by the Tillinghast Reflector Echelle Spectrograph (TRES). 
We then combine our derived metallicities with a kinematic analysis to identify confirmed planets in the thick disk that may have been previously overlooked in the literature. Our uniform dataset represents a key resource for studies of stellar populations, exoplanet demographics, and star-planet correlations. We use this catalog to explore some of these trends here.

This paper is structured as follows. In Section~\ref{sec:spectral_data}, we discuss the TRES observations and describe our sample selection. In Section~\ref{sec:specanalysis}, we describe {\tt uberMS}, the spectral code we use to derive our stellar parameters, and we outline our framework and methodology. In Section~\ref{sec:results}, we discuss our findings, and we present our conclusions in Section~\ref{conclusions}.

\section{Spectral Data} \label{sec:spectral_data}
\subsection{Observations: TRES} \label{subsec:TRES}

The Tillinghast Reflector Echelle Spectrograph (TRES) is a fiber-fed high-resolution spectrograph on the 1.5m Tillinghast Reflector at the Fred Lawrence Whipple Observatory on Mt. Hopkins, Arizona, USA \citep{Szentgyorgyi:2007}. It has a resolving power of $R = 44,000$ and covers the optical wavelength range between 3860 – 9100 {\AA}. 

TRES plays a key role in the TESS Follow-up Observing Program (TFOP\footnote{https://tess.mit.edu/followup/}; \citealt{Collins:2018}). It has been used to obtain reconnaissance spectra of thousands of TESS Objects of Interest (TOI) host stars, mostly with declinations north of $-20^{\circ}$ and going as faint as V = 13.5 mag. The first reconnaissance spectrum of each star is scheduled to be near one of the quadratures according to a photometric ephemeris based on the TESS transit times and assuming a circular orbit.  If the first observation reveals a spectrum suitable for radial-velocity determinations, a second reconnaissance spectrum is scheduled near the opposite quadrature.  This strategy is designed to give the most leverage for detecting orbital motion.  For those stars where the two spectra yield an offset larger than 50 or 100 m/s, additional observations are often undertaken to derive orbital solutions. Thus, the target stars analyzed in this paper almost always have two spectra and often have more than a dozen. This enormous database, which contains nearly 100,000 high-resolution spectra, presents a unique opportunity to perform a uniform spectroscopic characterization of the hundreds of stars with confirmed exoplanets contained within it. 

The spectra in the TRES archive used here were reduced using the pipeline from \cite{Buchhave:2010}, in which the the data are flat-fielded using exposures of tungsten filament continuum lamps and wavelength calibrated with Th-Ar hollow cathode lamps. In addition to these standard procedures, we further preprocess the spectra to ensure they are in the format expected by our spectral code, {\tt uberMS}. To do this, we follow the methodology by \cite{Pass:2025}, who tested the performance of TRES-{\tt uberMS} and determined the parameter uncertainties achievable with this framework. In particular, \cite{Pass:2025} analyzed hundreds of stars with TRES spectra with {\tt uberMS}, including from the Hyades cluster, the SPOCS catalog \citep{Valenti:2005, Brewer:2016}, and stars with constraints on $T_{\rm eff}$ and $\log g_{*}$ from interferometric radii \citep{Soubiran:2024}. Based on their performance tests on these samples, they found that TRES-{\tt uberMS} yields parameter errors for dwarfs of approximately 100~K in $T_{\rm eff}$, 0.09~dex in $\log g_*$, and 0.04~dex in [Fe/H] and 0.03 in [$\alpha$/Fe], with larger errors for mid-to-late K dwarfs, although the latter reflects fundamental limitations in stellar models and is not unique to this methodology. We note that we do not apply the empirical temperature-dependent [Fe/H] correction introduced in \cite{Pass:2025}, which accounts for offsets of up to $\sim$0.1 dex at $T_{\rm eff} \sim 4800$ K. The performance metrics quoted above therefore describe the native behavior of TRES–{\tt uberMS} and serve as a reference point for our analysis.

\begin{figure*}[ht!]
\begin{center}
    \includegraphics[width=0.8\textwidth]{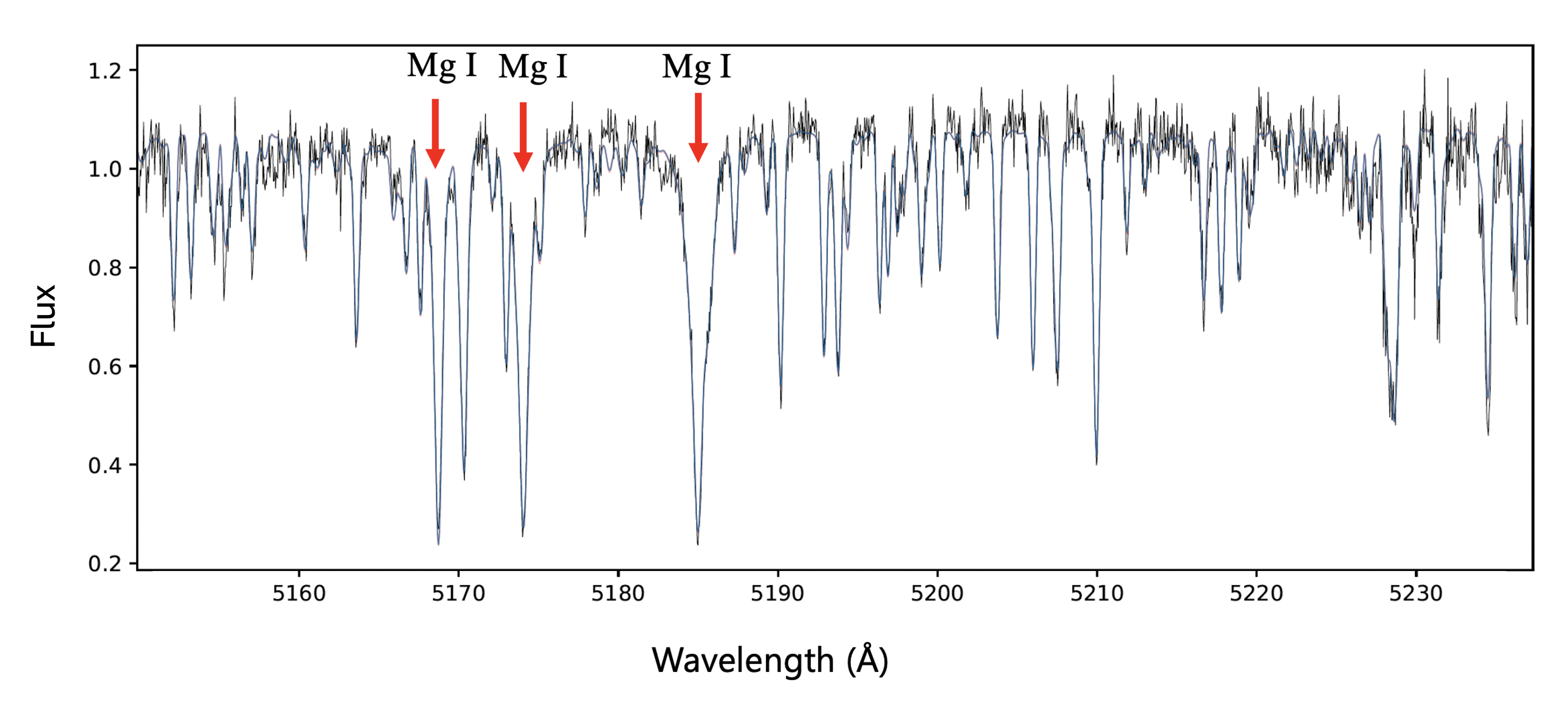}
\end{center}

\caption{Mg b triplet absorption features at 5167 {\AA}, 5172 {\AA}, and 5183 {\AA} in a representative TRES spectrum in our sample. These lines provide strong constraints on [$\alpha$/Fe].}
\label{fig:Mgtriplet} 
\end{figure*}

Since TRES-{\tt uberMS} is trained on spectra centered around the Mg b triplet (5150--5300 {\AA}; shown in Figure~\ref{fig:Mgtriplet}), we crop our spectra to this region, which corresponds to orders 23 and 24 of the TRES spectra,  and then normalize each order by the median flux. We cross-correlate each spectrum to shift them to a stellar rest frame and convert from air to vacuum wavelengths. We also perform a blaze correction based on nightly observations of Sirius and Vega, as described in detail in \cite{Pass:2025}. Some of the routines we implemented were directly adapted from the {\tt tres-tools} package\footnote{https://github.com/mdwarfgeek} written by Jonathan Irwin. Once these steps are applied, we co-add all observations of each star to increase the signal-to-noise ratio (SNR) of our master observed spectrum for each star. Our targets had on average 6 available spectra and a median SNR of 19.

\subsection{Sample Selection}  \label{sec:sample}

We began our sample selection by querying the NASA Exoplanet Archive\footnote{https://exoplanetarchive.ipac.caltech.edu/index.html} \citep{Christiansen:2025}; specifically, the Planetary Systems Composite Data table on November 2024 and compiled a list of all confirmed planet hosts in the literature. We then cross-matched this initial list by coordinates with the TRES database and identified 1082 exoplanet host stars with at least one archival TRES observation. In this search, we excluded stars with reported effective temperatures ($T_{\rm eff}$) below 4800 K, due to the inherent limitations in our stellar models and the increasing complexity of stellar spectra for cooler stars. This threshold is also based on the tests by \cite{Pass:2025}, who found that the performance of TRES-{\tt uberMS} is poorer for mid-to-late K dwarfs. We also excluded stars with $T_{\rm eff} \geq 6500$ K, since the performance of  TRES-{\tt uberMS} was tested mainly on benchmark samples with stellar temperatures below this value. In addition, we observed a worse performance for stars with $T_{\rm eff} \geq$ 6500 K, based on discrepancies in our results between the different modes of our spectral analysis code (see Sections~\ref{subsec:uberMS} and \ref{subsec:failure_modes}).

\section{Spectroscopic Analysis}  \label{sec:specanalysis}

\subsection{{\tt uberMS}} \label{subsec:uberMS}

To analyze our data, we use the spectral code {\tt uberMS}\footnote{\url{https://github.com/pacargile/uberMS}}, which determines stellar parameters using a Bayesian framework. A full description of the code will be presented in Cargile et al.\ in prep., but we briefly summarize its capabilities here.

The {\tt uberMS} program employs a neural network trained on grids of theoretical spectra and is based upon {\tt The Payne} \citep{Ting:2019}, named in honor of Cecilia Payne-Gaposhkin \citep{Payne1925}, and {\tt MINESweeper} \citep{Cargile:2020}.  {\tt uberMS} incorporates optical spectra and photometry to determine stellar parameters and posterior distributions within a hierarchical Bayesian model. Specifically, it calculates $T_{\rm eff}$, surface gravities, bulk metallicity [Fe/H] and $\alpha$-enrichment [$\alpha$/Fe], projected rotational velocities $v_{*}$sin$i$, stellar radii, and distances. The code interpolates 1D LTE stellar models using the {\tt ATLAS-12} model atmospheres \citep{Kurucz:1970,Kurucz:2005} and line lists provided by R.\ Kurucz. It uses the {\tt SYNTHE} radiative transfer code \citep{Kurucz:2003} and adopts the solar abundances from \cite{Grevesse:1998}. To constrain [$\alpha$/Fe], {\tt uberMS} is largely informed by the Mg triplet features at 5167 {\AA}, 5172 {\AA}, and 5183 {\AA}, and Fe I lines throughout the spectra.

{\tt uberMS} can additionally incorporate stellar isochrones to constrain stellar masses similar to the approach used by the {\tt MINESweeper} code. In particular, it uses the MESA Isochrones and Stellar Tracks (MIST; \citealp{Dotter:2016,Choi:2016}) to estimate these parameters. In turn, MIST uses the stellar evolution code MESA to compute stellar evolutionary tracks for stars with masses between 0.1 -- 300 $M_{\odot}$ and a wide range of metallicities ($-$4.0 $<$ [Z/H] $<0.5$) and ages (see \cite{Choi:2016} and \cite{Dotter:2016} for more details). Thus, {\tt uberMS} can function in two modes: one mode fits a stellar atmospheric model and stellar energy distribution (SED) to the spectrum and photometry (hereafter ``TP mode", after {\tt The Payne}). In the second mode (hereafter ``MS mode", after {\tt MINESweeper}), it does that and simultaneously fits the MIST isochrones. We note that the MS mode results are preferred, as the simultaneous fitting of stellar evolutionary models provides additional constraints and leads to smaller parameter uncertainties. 

\begin{figure*}[!ht]
\begin{center}
    \includegraphics[width=1.0\textwidth]{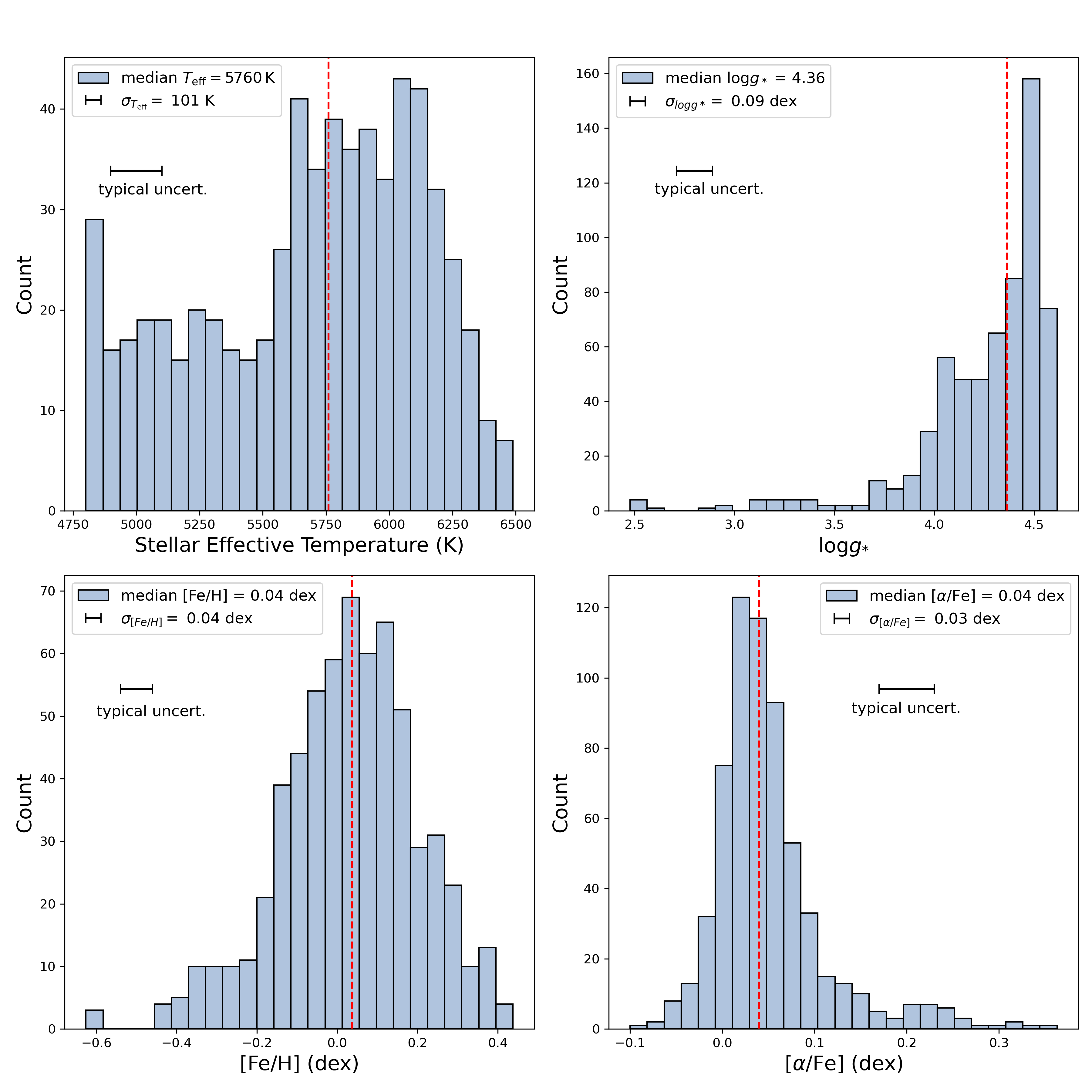}
\end{center}
\caption{Distribution of effective temperatures ($T_{\rm eff}$), surface gravities ($\log g_*$),  [Fe/H], and [$\alpha$/Fe] for our full, final sample. The red dashed vertical lines denote the median of each distribution, and the error bars represent the typical (median) uncertainty in each parameter.}
\label{fig:TP_hist}
\end{figure*}

\subsection{Analysis} \label{sec:analysis}

Once all our spectra are preprocessed, co-added, and in the format expected by {\tt uberMS} as described in Section~\ref{subsec:TRES}, we gather all available information for each target to use as starting points and priors in our analysis. For every star, orders 23 and 24 are jointly fit, with various spectroscopic parameters allowed to vary within each order. To better constrain $T_{\rm eff}$ and visual extinction, we fit the spectral energy distribution of each star based on available photometry from \textit{Gaia} $G$, $G_{BP}$, and $G_{RP}$ \citep{GaiaDR3}, and 2MASS $J$, $H$, and $K$ magnitudes \citep{Cutri:2003}.

Following the methodology of \cite{Pass:2025}, we initialized each fit with starting values of: $T_{\rm eff}=5000$~K, $\log g = 4.5$~dex, [Fe/H]$=0.0$~dex, [$\alpha$/Fe]$=0.0$~dex, $\log R=0.0$~dex, and $v_*=2$~kms$^{-1}$ in the TP mode. We used parallaxes from {\it Gaia} DR3 \citep{Gaia:2023}, and constrained the visual extinction along the line of sight $A_{V}$, using the reddening values from the all-sky, three-dimensional dust maps from \cite{Vergely:2022}. We queried these values using the G-Tomo ESA database\footnote{\url{https://explore-platform.eu/articles/gtomo_datalabs}}. We uniformly bound [Fe/H] and [$\alpha$/Fe] for all our stars between [$-$4, 0.5] and [$-$0.2, 0.6] dex, respectively. We placed a prior on the  microturbulence based on the empirical relationship between $T_{\rm eff}$ and $\log g$ from \cite{Bruntt:2012}. We fixed the spectroscopic jitter to 0.015 in units of normalized flux. This choice was set based on the tests by \cite{Pass:2025}, who found that allowing this parameter to float yields inconsistent results for multiple observations of the same star. For the analysis in the MS mode, i.e., employing isochrones, the fit is initialized with an Equivalent Evolutionary Phase (EEP) at EEP = 250 and is bound between 200 and 600. Our EEP prior confines the fit to evolutionary stages between the zero-age main sequence and the tip of the red giant branch, with greater weight assigned to main-sequence solutions. In addition, we applied a prior on the latent age parameter to penalize solutions older than the age of the Universe. We also set the initial [Fe/H] and [$\alpha$/Fe] at 0, and set an initial mass of 1$M_{\odot}$. Our full set of initial values and priors are listed in Table~\ref{tab:priors}. 
%not including this since we do not report masses 
%Finally, we bound the stellar mass between 0.5 and 2 $M_{\odot}$

% \begin{figure*}
%     \centering
%     \includegraphics[width=0.45\textwidth]{radius.png} &
% \includegraphics[width=0.45\textwidth]{luminosity.png}\\
%     \caption{Comparison between stellar parameters (stellar radius: left panel; luminosity: right) from the GBS3 benchmark sample of stars with interferometric radii \citep{Soubiran:2024} and our parameters derived from TRES spectra using {\tt uberMS}. The top plots show the direct comparison for dwarfs (in blue) and evolved stars (in red); a 1:1 line is plotted for reference. The bottom plots show the percent difference in each property, with the colored shaded regions representing the individual RMS.} 
%     \label{fig:RadLumtests}
% \end{figure*}

\begin{figure*}
\centering
\includegraphics[width=0.49\textwidth]{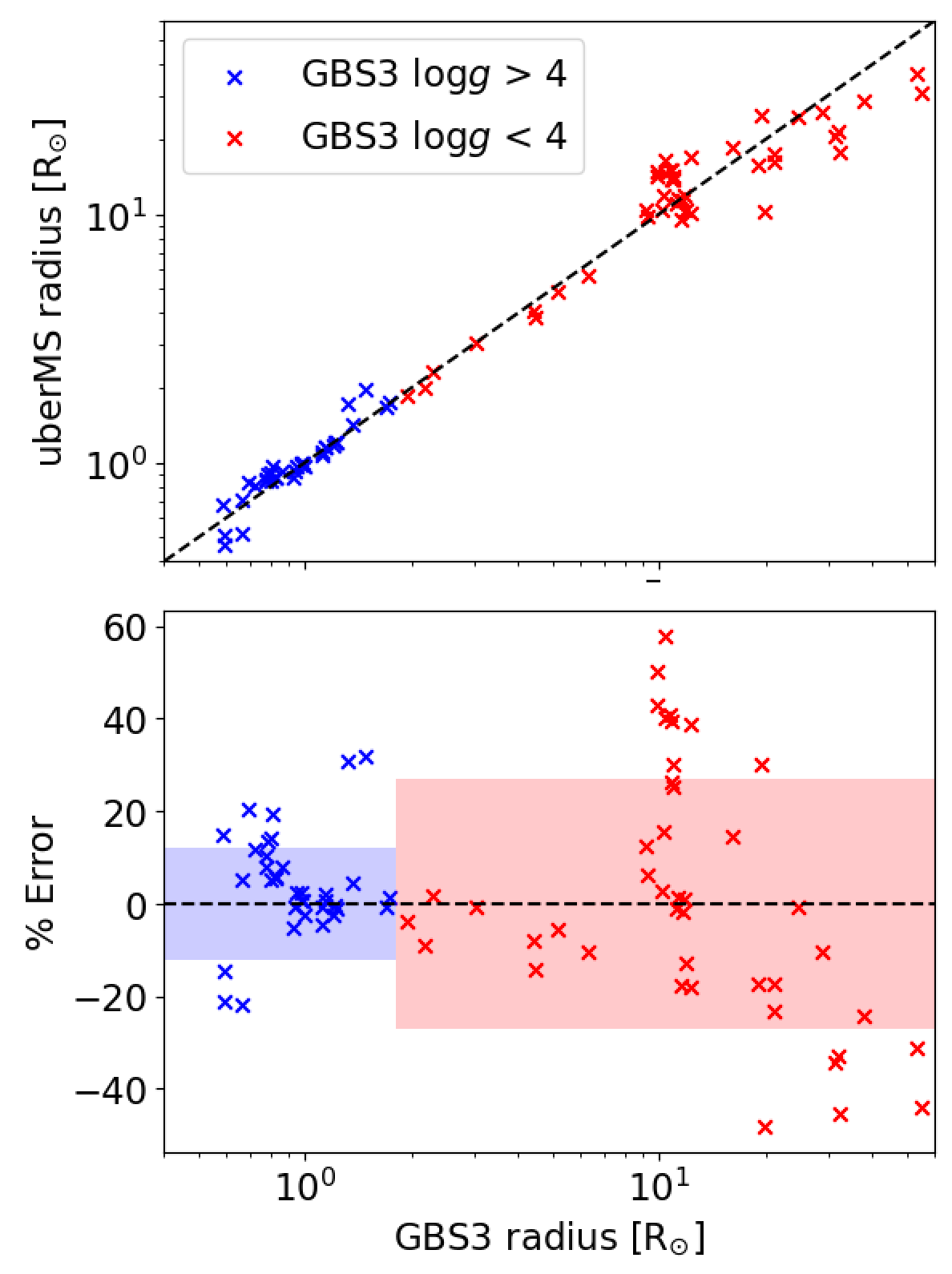}
\hfill
\includegraphics[width=0.49\textwidth]{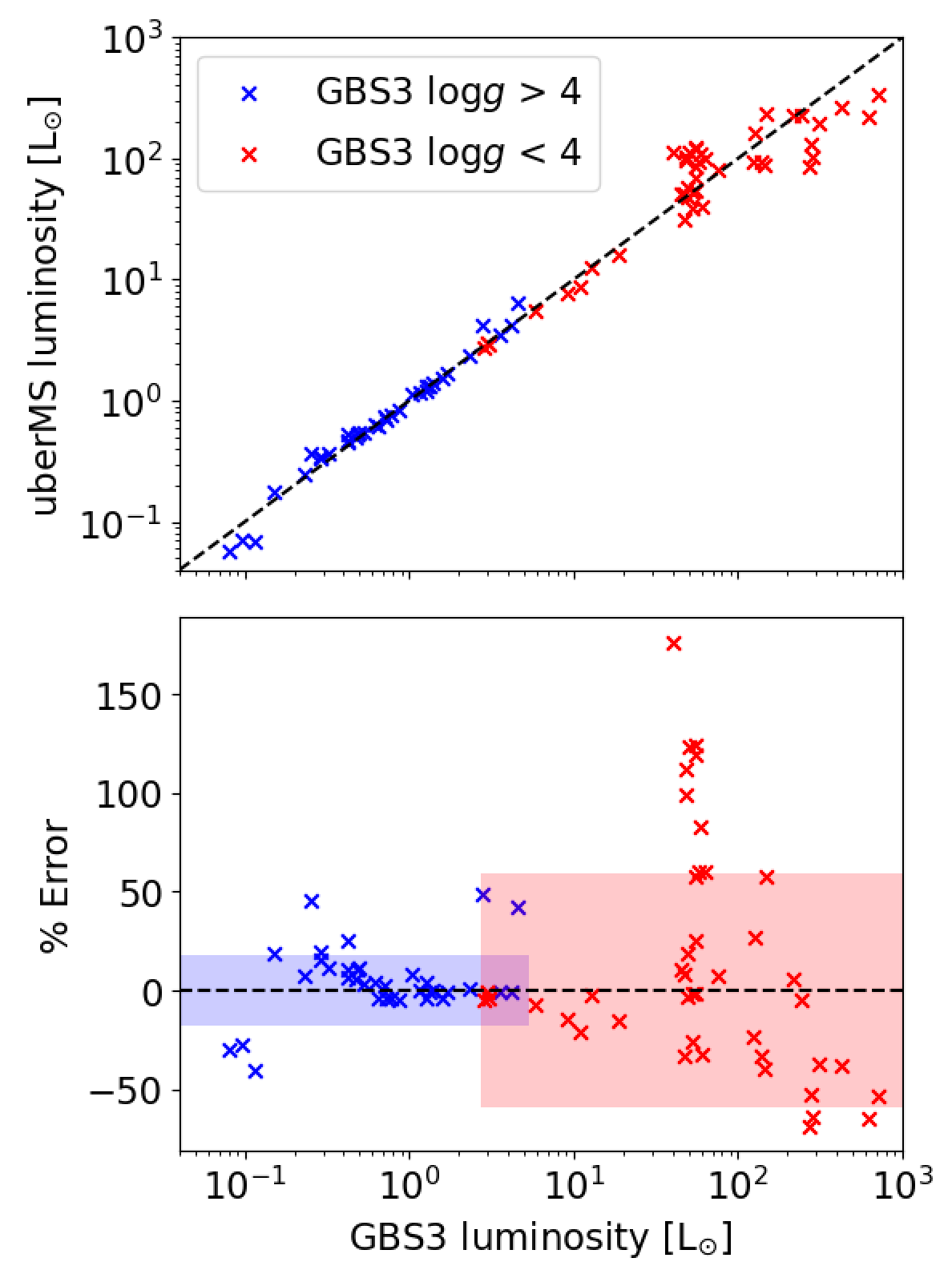}
\caption{Comparison between stellar parameters (stellar radius: left panel; luminosity: right) from the GBS3 benchmark sample of stars with interferometric radii \citep{Soubiran:2024} and our parameters derived from TRES spectra using {\tt uberMS}. The top plots show the direct comparison for dwarfs (in blue) and evolved stars (in red); a 1:1 line is plotted for reference. The bottom plots show the percent difference in each property, with the colored shaded regions representing the individual RMS.}
\label{fig:RadLumtests}
\end{figure*}

\begin{figure}[!ht]
\vspace{.1in}
\centering
\includegraphics[width=1\linewidth]{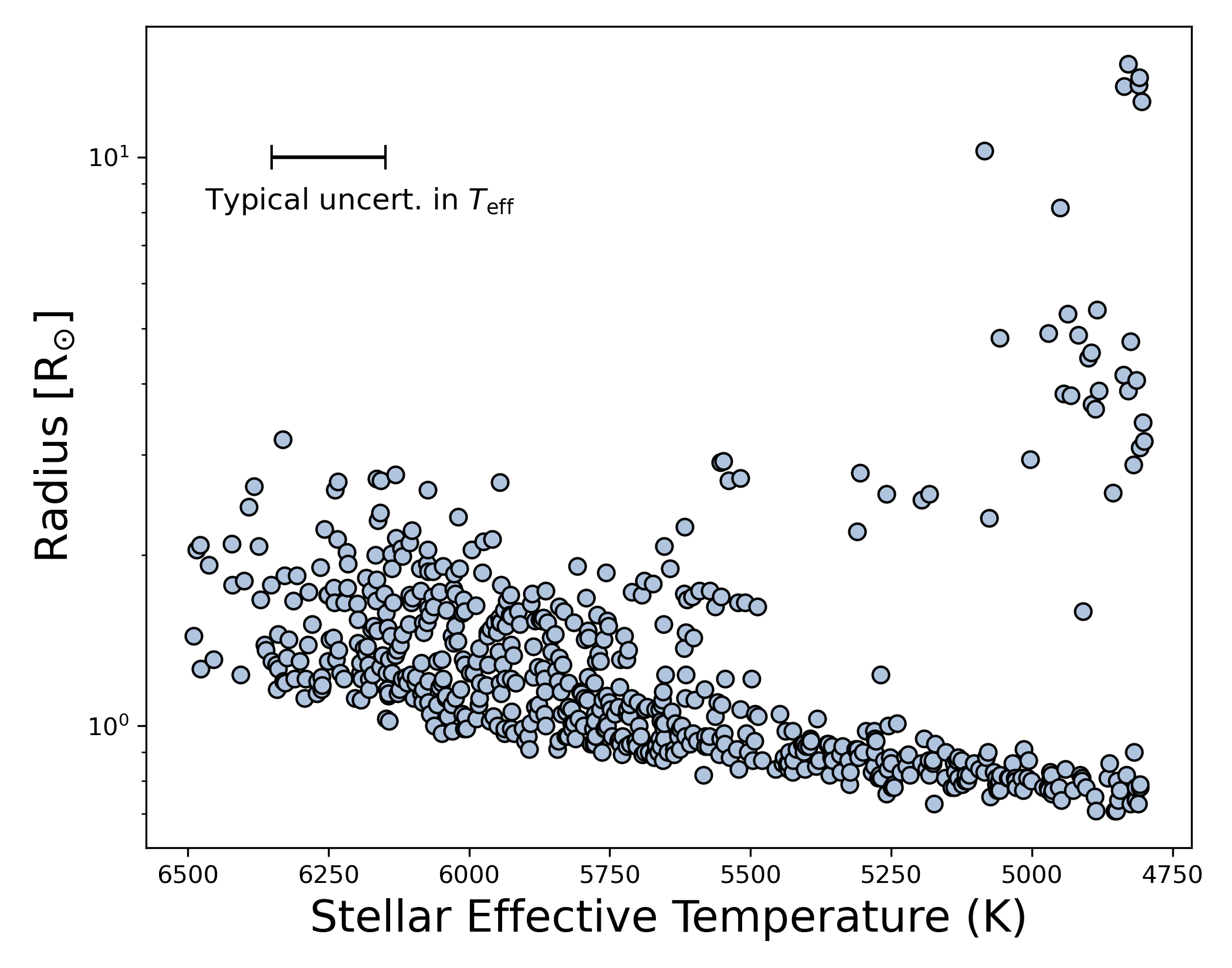}
\caption{Stellar radius and effective temperature distribution of our full sample of planet-hosting stars observed by TRES.}
\label{fig:HR_diagram} 
\end{figure}

\section{Results and Discussion} \label{sec:results}

After running {\tt uberMS} on our full sample, we visually inspected the results to evaluate the quality of the fits and check that the derived stellar parameters and posterior distributions were physically reasonable. Specifically, we examined the standard diagnostic plots from {\tt uberMS}, which include the spectra with the best-fit models and residuals, the spectral energy distribution fit, and the posterior distributions (see Figure 1 of \citealt{Cargile:2020}). We rejected any stars that exhibited large or structured residuals in the spectral fits, clear mismatches in temperature- or gravity-sensitive features, or SED fits inconsistent with the inferred stellar parameters. Stars with SNR $\lessapprox$ 10 were generally found to yield unreliable parameter estimates and were excluded, particularly when accompanied by poor spectral or SED fits. We also removed stars with derived $v_{*}$sin$i$ $>$ 35 km/s, since beyond this value, significant rotational broadening in the spectral features makes the metallicity measurements unreliable \citep{Torres:2012}. After implementing these quality cuts, we ended up with a final sample of 625 planet hosts stars out of an initial sample of 1082.

In the MS mode, i.e., including isochrones, we obtain median distribution values of $T_{\rm eff}$ = 5760 K, $\log g_*$ = 4.36 dex, [Fe/H] = 0.04 dex, and [$\alpha$/Fe] = 0.04 dex. We show histograms of the median of the posterior distributions for these parameters in Figure~\ref{fig:TP_hist}. Note that the slightly supersolar median metallicities reflect the fact that these stars are all planet hosts.

While running {\tt uberMS} in both modes produced consistent results for the majority of the sample, we recommend using the parameters from the MS mode (with MIST isochrones). The inclusion of isochrones provides a physically motivated prior on the stellar evolutionary stage, enabling tighter and more realistic constraints on the surface gravity ($\log g_{*}$) and stellar radius. This is especially critical for main sequence and subgiant stars, for which $\log g_{*}$ is notoriously difficult to measure spectroscopically. The formal precision for our stellar parameters from {\tt uberMS} for dwarf stars in our analysis are 100~K in $T_{\rm eff}$, 0.09~dex in $\log g_*$, 0.04~dex in [Fe/H], and 0.03~dex in [$\alpha$/Fe], while for evolved stars they are 100~K, 0.14~dex, 0.10~dex, and 0.15~dex, respectively \citep{Pass:2025}.

To evaluate the systematic errors on the stellar radius and luminosity from TRES-{\tt uberMS}, we tested our methodology on a sample of benchmark stars with interferometrically derived radii, complementing the tests performed by \cite{Pass:2025}. We ran {\tt uberMS} on 85 stars with available TRES data from the {\it Gaia} FGK Benchmark Sample (GBS3; \citealt{Soubiran:2024}), which derived $T_{\rm eff}$ and $\log g_*$ for a set of dwarfs and evolved stars with precise angular diameters and parallaxes. These interferometric radii are much more precise than spectrophotometrically determined radii, and therefore they can be used to determine the systematic errors in our measurements. We obtained 12\% and 27\% radius RMS errors for dwarfs ($\log g_* > 4$ dex) and evolved stars ($\log g_* < 4$ dex), respectively. While the stellar luminosity is not a direct output of {\tt uberMS}, we calculate it via
\begin{equation}
    \frac{L}{L_{\odot}} = \bigg(\frac{R}{R_{\odot}}\bigg)^2 \bigg(\frac{T_{\rm eff}}{T_{\odot}}\bigg)^4.
\end{equation} 
\noindent Propagating the errors from the fundamental quantities we obtain from {\tt uberMS}, we determined 18\% and 59\% luminosity RMS errors for dwarfs and evolved stars, respectively. Figure~\ref{fig:RadLumtests} illustrates both parameter comparisons.

Our final results and stellar parameters are presented in Table~\ref{tab:stellarprops}. We compared our results to the default stellar parameters reported in the NASA Exoplanet Archive, and we found generally good agreement, with mean offsets of $\Delta T_{\rm eff} = -28$ K in $T_{\rm eff}$ and $\Delta[\mathrm{Fe/H}] = -0.028$ dex in [Fe/H], and RMS scatters of 146 K and 0.11 dex, respectively, where $\Delta$ is defined as this work minus literature.

\begin{figure*}[ht!]
\begin{center}
    \includegraphics[width=0.8\textwidth]{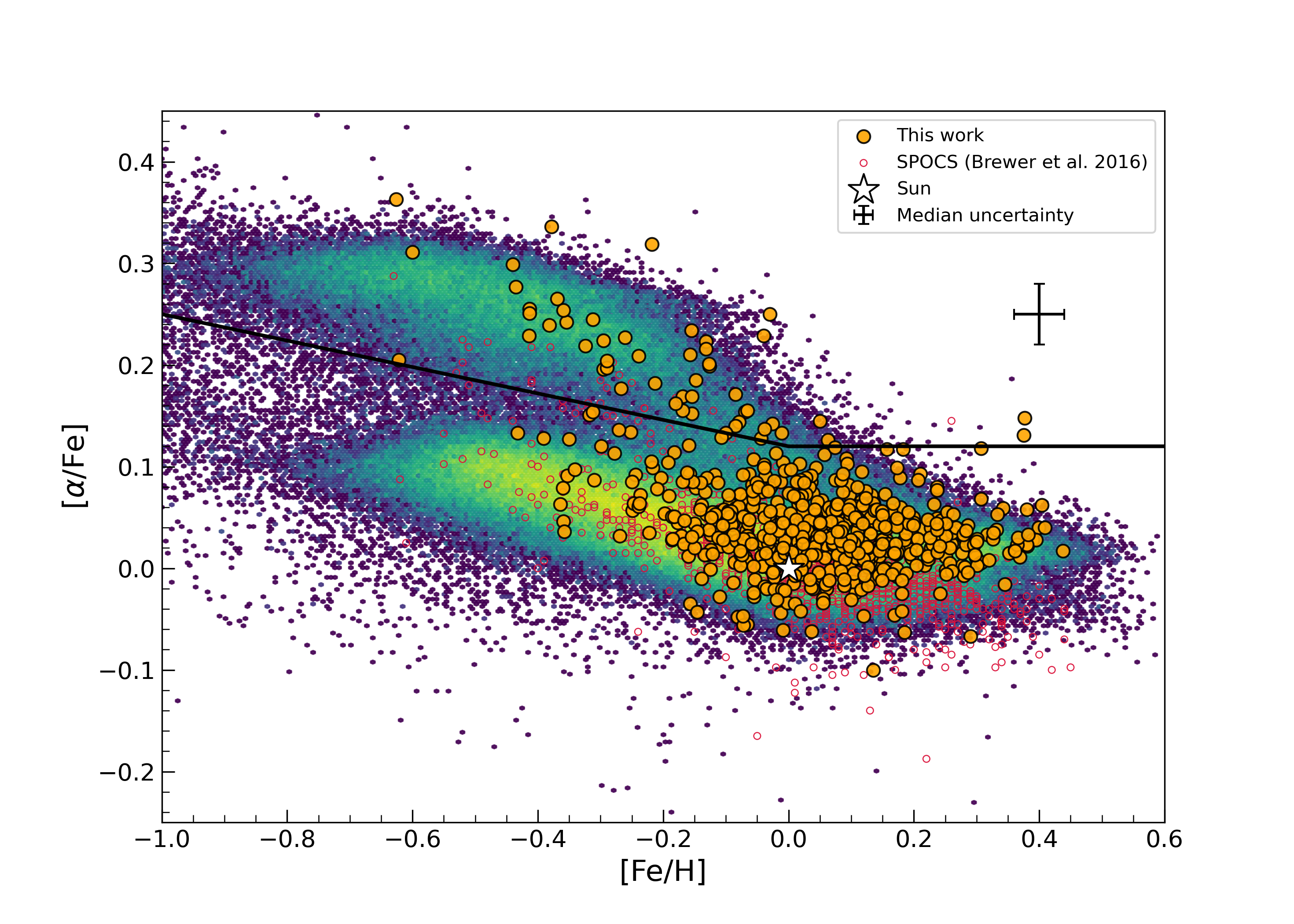}
\end{center}

\caption{Stellar [$\alpha$/Fe] versus [Fe/H] for planet-hosting stars analyzed in this work (orange circles). The heat map represents the density of stars from APOGEE DR17 in the [$\alpha$/M] vs [Fe/H] parameter space for reference. The black line approximately divides the thin and thick disk populations (below and above the line, respectively) as defined by \citealt{Weinberg:2019}. The red points represent stars from the SPOCS sample from \cite{Brewer:2016}. The Sun is over-plotted for reference.}
\label{fig:alpha_FevsFe} 
\end{figure*}

\subsection{Failure Modes and Stellar Binaries}\label{subsec:failure_modes}

As a sanity check, we compared our results with and without isochrones, which should, in principle, yield consistent results. Therefore, any significant discrepancies between these modes could be indicative of either problematic fits and/or unresolved stellar companions. Since our stellar parameters are not reliable for such cases, we removed stars for which the differences in $T_{\rm eff}$ between the TP and MS modes exceeded the nominal  1$\sigma$ uncertainty of 100~K (60 stars in total).

\subsection{Evolved Stars}

The final sample studied here covers a wide range of stars across the H-R diagram, and we identified 87 likely evolved or giant stars with $\log g_*$ $< $ 4, which can be seen in the upper right part of the temperature-radius distribution of our sample in Figure~\ref{fig:HR_diagram}. We note, however, that parameter uncertainties are larger for these stars compared to dwarfs.

\subsection{Brown Dwarfs}

In addition to these evolved stars, there are at least 4 stars with brown dwarf companions in our sample, which we define as those with masses $M \geq 13 M_{J}$. These are TIC 248048656, TIC 9815387 (nu Oph b and c), TIC 251662732, and TIC 20096620. Constraining the atmospheric composition of brown dwarfs can yield key insights into their formation and into the atmospheres and formation of giant planets. One path to constraining their atmospheric properties is through abundance measurements of Sun-like stellar primaries in systems with brown dwarf companions, since we can assume that both components have similar metallicities (e.g., \citealt{Gonzales:2020, Phillips:2024}). However, Sun-like stars with brown dwarf companions are rare, and the available abundances for such systems are often riddled with systematics arising from disparate measurements in the literature. Thus, the uniform metallicities of brown dwarf stellar hosts presented here may provide useful insights into the formation histories of these systems. Future constraints on C/O ratios of the hosts in comparison to the brown dwarf can further help disentangle formation pathways \citep{Phillips:2024}, although such work is beyond the scope of this paper.

\subsection{Alpha-enrichment and Galactic Kinematics} \label{subsec:alpha}

\begin{figure*}[!ht]
\begin{center}    \includegraphics[width=0.80\textwidth]{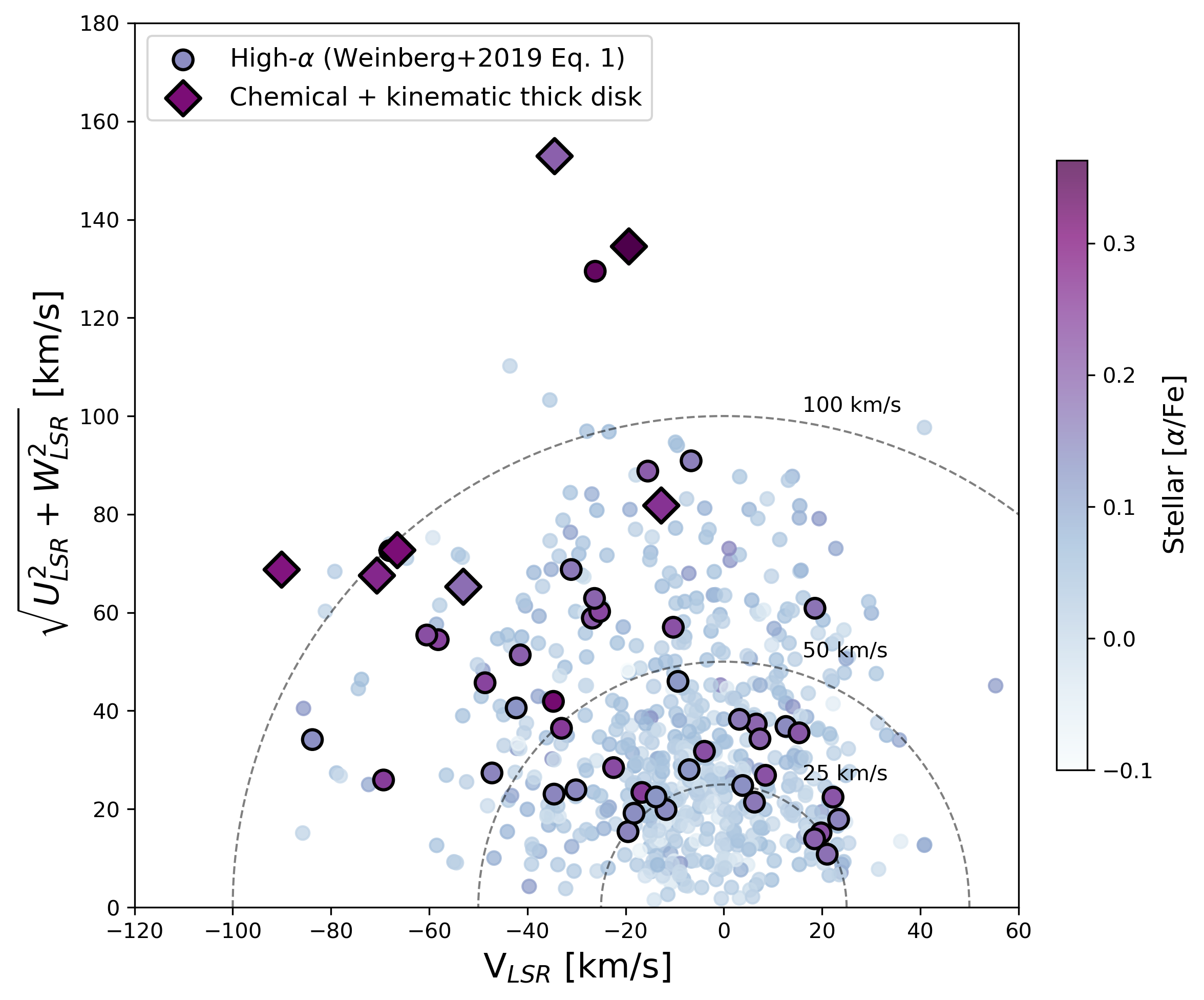}
\end{center}
\caption{Toomre diagram of the stars in our sample. The concentric dashed circles represent constant total space velocities. The points are color-coded by stellar [$\alpha$/Fe] enrichment, and stars that are chemically consistent with being in the thick disk (above the dividing thick/thin disk line in \citealt{Weinberg:2019}) are highlighted with black edges. Stars that are both kinematically and chemically consistent with the thick disk are plotted as diamonds.}
\label{fig:Toomre}
\end{figure*}

To identify stars in our sample that may be chemically associated with the galactic thick disk, we plot [$\alpha$/Fe] versus [Fe/H] in Figure~\ref{fig:alpha_FevsFe}. Stars from our work are denoted in orange, and stars from APOGEE DR17 and from the SPOCS sample \citep{Brewer:2016} are additionally plotted for reference. The black line denotes the approximate chemical boundary between the thin and the thick disk as defined by \cite{Weinberg:2019} (their Eq.\ 1, substituting [$\alpha$/Fe] for
[Mg/Fe]). If we adopt this chemically-motivated definition of the thick disk, and using as reference the line delineated by \cite{Weinberg:2019}, there are 51 stars in our sample consistent with the thick-disk population.

To further assess thick-disk membership, we complemented our chemical analysis with galactic kinematics and calculated the probability that each star belongs in the thick disk following a kinematic definition. We first cross-matched our sample with {\it Gaia} DR3, and using the parallax, proper motions, and absolute systemic radial velocities given from that database, we determined the U,V,W heliocentric space velocities of each star. We then converted these velocities to the 3-dimensional space motion relative to the local standard of rest ($\rm U_{LSR}$, $\rm V_{LSR}$,$\rm W_{LSR}$), using the peculiar velocities for the Sun from \cite{Schoenrich:2010}. Finally, we computed the probability of a star being in the thick disk, divided by the probability of being in the thin disk $P (\rm thick)/ P(\rm thin)$ as defined by \cite{Bensby:2003}. This work estimated thick disk membership from the U,V,W space velocities, and the velocity dispersions ($\sigma_{U}$, $\sigma_{V}$, $\sigma_{W}$), and asymmetric drift ($V_{\rm asym}$) of each stellar population (thin disk, thick disk, and halo). We adopted the $\sigma_{U}$, $\sigma_{V}$, $\sigma_{W}$ and $V_{\rm asym}$ values from Table 1 of \cite{Bensby:2003} and we assumed stellar population fractions of 83\% for the thin disk, 17\% for the thick disk, and 0.1\% for the halo \hbox{\citep{Fantin:2019}.} Following the classification scheme of \cite{Bensby:2003}, we defined stars as kinematic thick disk members if $P (\rm thick)/ P(\rm thin)$ $\geq$ 10. We found 14 stars in our sample above this threshold. In Figure~\ref{fig:Toomre}, we present our sample in a Toomre diagram, with points color-coded by our derived [$\alpha$/Fe], and we highlight stars that are chemically consistent with thick-disk membership (i.e., above the thick-disk boundary in \citealt{Weinberg:2019}) as circles with black edges. Some of these stars also have relatively higher total space velocities consistent with the thick disk (70 $\leq$ $V_{\rm tot}$ $\leq$ 200 km/s). We found 7 stars that overlap in both samples, i.e., are both chemically and kinematically consistent with the thick-disk. These are: HD 233832, K2-111, K2-408, Kepler-511, Kepler-517, K2-180, and K2-337. We list all our kinematic and chemical thick-disk candidates in Table~\ref{tab:thickdisk}. We also included the number of $\sigma$ by which each candidate lies above thick-disk line (given by \citealt{Weinberg:2019}) in our final table. This quantity can be used as a rough likelihood of thick-disk membership, with higher values representing higher probabilities. Candidates close to the boundary, particularly evolved stars with larger abundance uncertainties, should be interpreted as less secure thick disk candidates.

Several of these stars had already been suspected or validated as thick disk stars, including TOI-561 (TIC 377064495; \citealt{Lacedelli:2020, Weiss:2021}), K2-180 (TIC 366411016; \citealt{Mayo:2018, Korth:2021}), HD 20329  (TIC 333657795; \citealt{Murgas:2022}), and HD 233832 (TIC 445811628; \citealt{Barbato:2019}). For K2-111, \hbox{\cite{Fridlund:2017}} calculated an [Fe/H] = $-$0.53 $\pm$ 0.05 dex, an age of $\sim$10 Gyr, and kinematics consistent with the thick disk. We calculate a high $\alpha$-enrichment of [$\alpha$/Fe] $=$ 0.26 $\pm$ 0.02 dex, and a high thick/thin disk probability ratio  $P (\rm thick)/ P(\rm thin)$ of 16, which provides further evidence that K2-111 indeed resides in the thick disk.

\subsection{Correlations Between Stellar Chemistry and Exoplanet Properties} \label{subsec:metallicity_correlations}

\subsubsection{[$\alpha$/Fe] and Planet Multiplicity} \label{subsec:metallicity_multiplicity}

\begin{figure}[ht!]
\vspace{.1in}
\centering
 \includegraphics[width=1\linewidth]{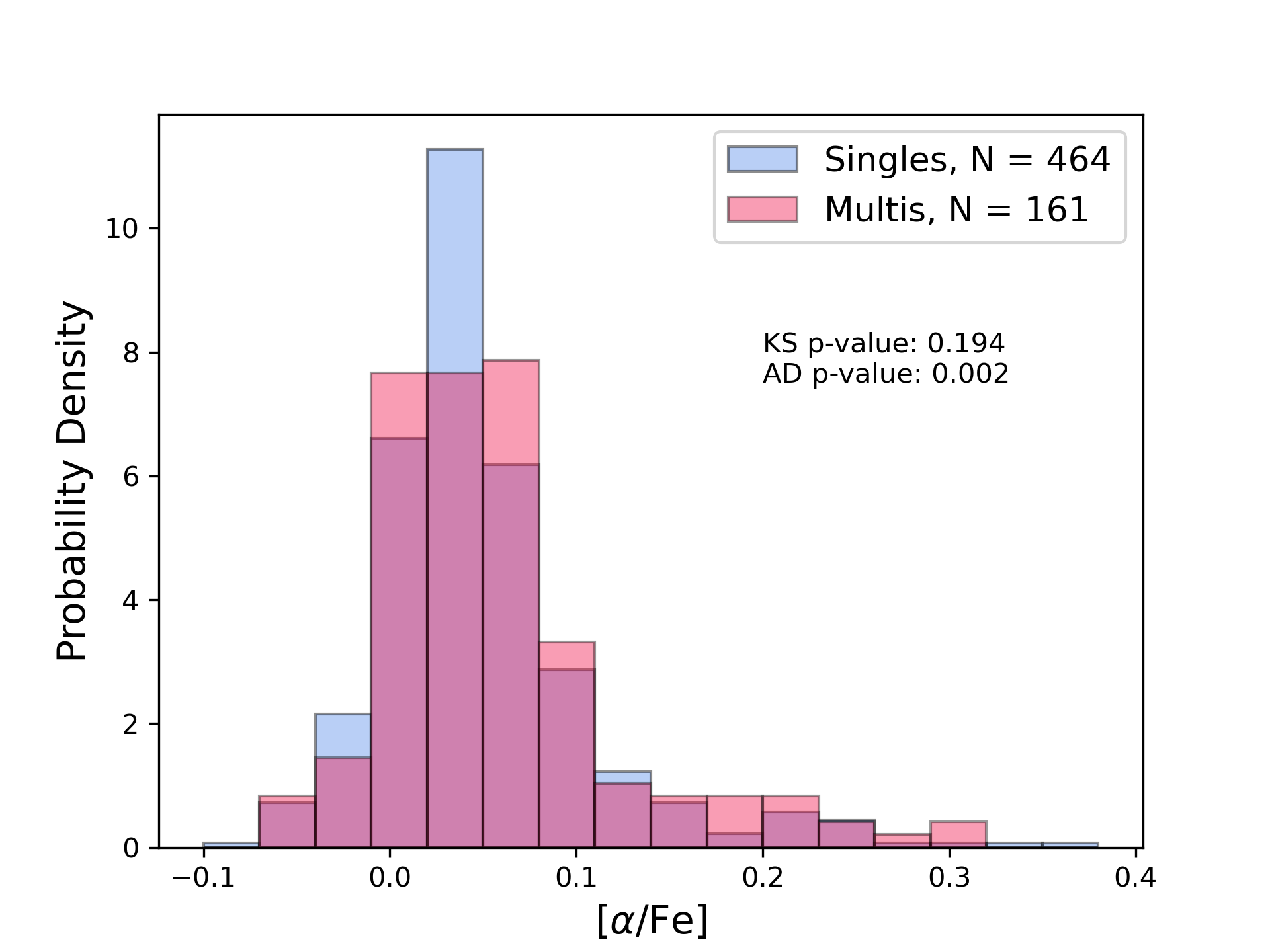}
%\vspace{-.25in}
\caption{Normalized [$\alpha$/Fe] distributions of stars hosting a single known planet (shown in blue) and those hosting multiple planets (in red).}
\label{fig:metal_multplicity} 
\end{figure}

We investigate the relationship between [$\alpha$/Fe] and planet multiplicity. Many authors have previously investigated the differences between [Fe/H] bulk metallicity and planet multiplicity \citep[e.g.,][]{Weiss:2018, Anderson:2021, Munoz-Romero:2018, RodriguezMartinez:2023b}. However, no previous study has directly compared the [$\alpha$/Fe] distributions of single- and multi-planet host stars. We present here the first such analysis using our homogeneously derived [$\alpha$/Fe]. We first defined our stellar population as either a ``single" (with 1 confirmed planet) or a multi-planet host (2 or more confirmed planets based on the number of planets reported in the NASA Exoplanet Archive, specifically the {\tt sy\_pnum} column). There were 464 singles and 161 multis in our sample. We quantified the differences in their [$\alpha$/Fe] distributions with Kolmogorov-Smirnov (KS) and Anderson-Darling (AD) tests. The AD test detects a statistically significant difference ($p =$0.002), largely driven by differences in the distribution tails, whereas the KS test yields $p=$0.194, consistent with similar central distributions (see Figure~\ref{fig:metal_multplicity}). Thus, while we do not find strong evidence for a difference in the [$\alpha$/Fe] distributions of singles and multis, the excess of multi-planet systems at high [$\alpha$/Fe] may suggest that $\alpha$-enhanced stars may preferentially host compact multi-planet systems.

\subsubsection{[$\alpha$/Fe] Distribution of Giant-Planet Hosts} \label{subsec:alpha_giants}

\begin{figure*}[!ht]
\begin{center}
    \includegraphics[width=0.65\textwidth]{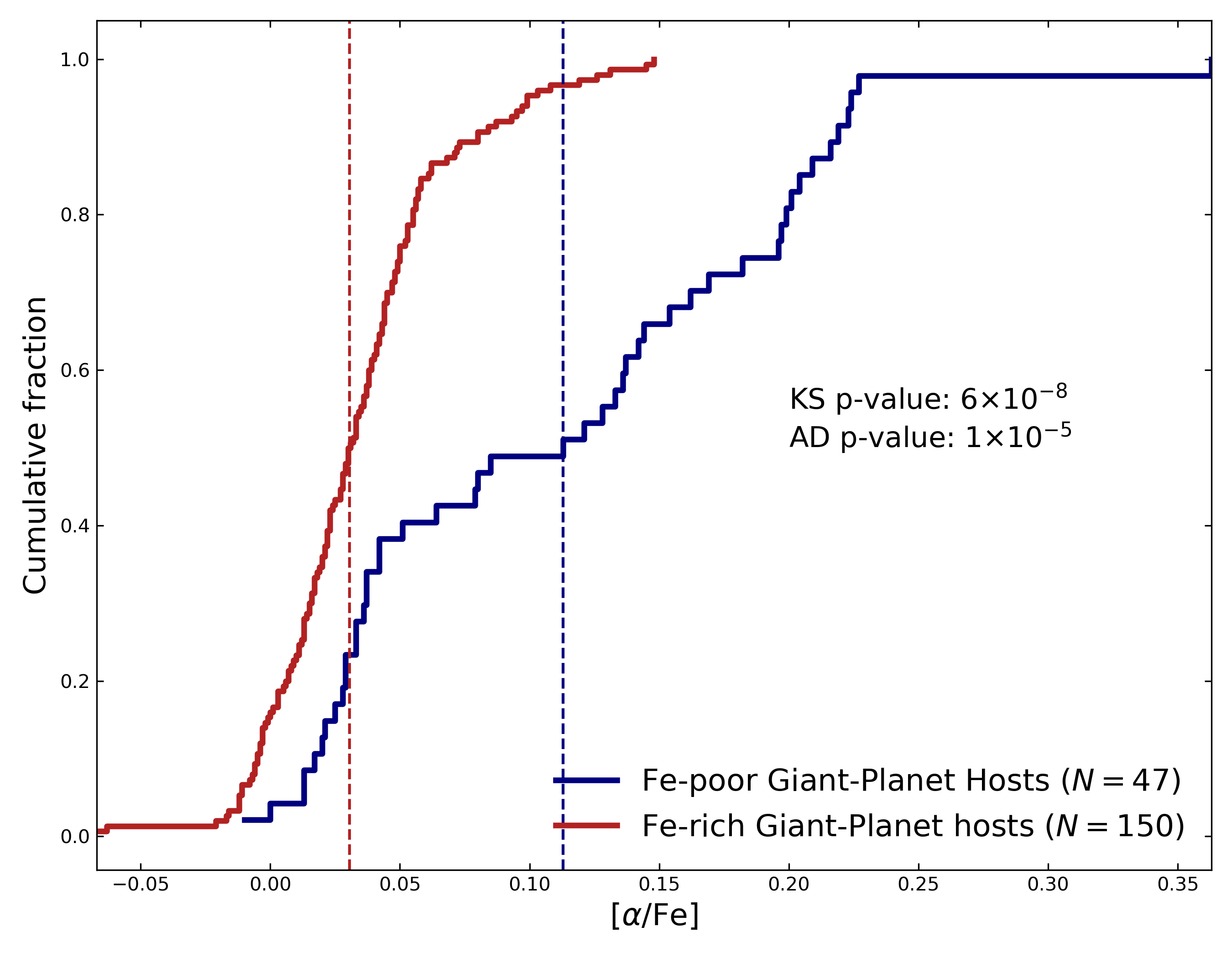}
\end{center}
\caption{Empirical cumulative distribution functions of [$\alpha$/Fe] for giant planet hosts with [Fe/H] $\leq$ 0 (Fe-poor; navy line), and for giant-planet hosts with [Fe/H] $>$ 0, in brown. The vertical dashed lines denote the median of each distributions. At low [Fe/H], giant-planet hosts have significantly elevated [$\alpha$/Fe] compared to metal-rich hosts. The KS and AD $p$-values are also plotted.}
\label{fig:GPs_CDFs}
\end{figure*}

The homogeneous chemical abundances presented in this catalog allow us to explore subtle questions regarding the role of $\alpha$-element enhancement in planet formation. While the correlation between giant-planet occurrence and stellar metallicity has been firmly established for nearly two decades, the influence of [$\alpha$/Fe] on planet demographics has only recently begun to be more deeply investigated as metallicities for planet hosts have become more widely available. In particular, it has been suggested that at low [Fe/H], $\alpha$-element enrichment may partially compensate for reduced iron content in the protoplanetary disk. For example, \citet{Haywood:2008} reported that 13 of the 14 giant planets known at the time around metal-poor stars ($-0.7 < [{\rm Fe/H}] < -0.2$) orbited $\alpha$-enhanced hosts, hinting that higher [$\alpha$/Fe] might help enable giant-planet formation in iron-depleted environments. We test this idea with our larger and homogeneously analyzed sample by comparing the [$\alpha$/Fe] distribution of giant-planet hosts with sub-solar metallicities ([Fe/H] $\leq$ 0; $N=47$) to that of metal-rich giant-planet hosts with [Fe/H] $>0$ ($N=150$). We classify giant planets as those with $R_{p} \geq 8R_\oplus$ and $M_{p} \leq 13 M_{J}$ (i.e., below the brown-dwarf boundary). We find that the Fe-poor giant-planet hosts exhibit  significantly higher median [$\alpha$/Fe] values (0.11) than the metal-rich giant-planet host population (0.03). This difference is highly statistically significant, with an AD $p$-value of $p = 1\times10^{-5}$, and a KS test $p$-value of $p = 6\times10^{-8}$. Figure~\ref{fig:GPs_CDFs} shows the empirical cumulative functions of both distributions. For completeness, we also compared the metal-poor giant-host subsample to the overall giant-host median [$\alpha$/Fe], which yields similarly significant differences (AD: $p = 1\times10^{-5}$; KS: $p = 5\times10^{-5}$), consistent with the increasingly $\alpha$-poor composition of the metal-rich thin disk.

We note that, in general, low-metallicity stars are on average more $\alpha$-enhanced than high-metallicity stars, regardless of planet occurrence. Therefore, to test whether the $\alpha$-enhancement of metal-poor giant-planet hosts we observe simply reflects the underlying Galactic [$\alpha$/Fe]–[Fe/H] relation, we compared them to metal-poor field stars from APOGEE DR17. We performed a bootstrap analysis in which we randomly drew $\rm N_{GP}$ stars from APOGEE with [Fe/H] $<$0, where $\rm N_{GP}$ is the number of stars in our Fe-poor giant-planet host sample, and computed the median [$\alpha$/Fe] of each draw. Repeating this process 10,000 times, we found that only $\sim$1.5\% of the APOGEE samples had median [$\alpha$/Fe] values exceeding the median value of the Fe-poor giant-planet hosts. This suggests that metal-poor giant-planet hosts are unusually $\alpha$-enhanced relative to typical metal-poor field stars, supporting the interpretation that $\alpha$-element enrichment may help compensate for lower iron abundances in giant-planet formation.

We also compared the [$\alpha$/Fe] distributions of giant-planet hosts to those hosting smaller planets ($R_{p} < 8R_\oplus$) and
we find no strong difference in the median [$\alpha$/Fe] values of giant-only and small-planet-only hosts (medians of 0.037 and 0.043, respectively), although the Anderson-Darling test suggests possible differences in the detailed distribution shapes (AD: $p$ = 0.000967, KS: $p$= 0.17). We performed an analogous test for small-planet hosts ($R \leq 4R_{\oplus}$. Splitting the sample by metallicity, we find that sub-solar small-planet hosts exhibit higher [$\alpha$/Fe] than their super-solar counterparts (medians of 0.06 vs 0.03). The difference is statistically significant, with KS: $p = 2\times10^{-8}$ and  AD: $p = 1\times10^{-5}$. Furthermore, these differences are unlikely to be driven solely by observational selection effects: below a given [Fe/H] threshold, there is no obvious detection advantage for planets orbiting $\alpha$-enhanced thick-disk stars compared to $\alpha$-depleted stars.

These results are consistent with the findings of \cite{Haywood:2008}, \cite{Adibekyan:2012} and more recently, \cite{Sharma:2024}, who also found an overabundance of Mg and Si for planet hosts at low-metallicity regimes. These results collectively support the idea that $\alpha$-element enrichment may partially compensate for a reduced inventory of iron, thereby enhancing planet formation across a broad range of planet sizes. More broadly, our findings suggest that [$\alpha$/Fe] may play an important role in planet formation and that planet demographics depend on detailed stellar chemistry and Galactic population context.

\subsubsection{[Fe/H], Period and Eccentricity} \label{subsec:metallicity_period_eccentricity}

We  performed a cursory investigation of potential correlations between host star metallicity, planet eccentricity, and orbital period. In particular, we revisited the canonical plot from \cite{Winn:2015} (Figure 3 of that paper), who showed that the most eccentric planets tend to orbit metal-rich ([Fe/H] $>$ 0) stars. This trend has been interpreted as evidence of stronger dynamical interactions in more massive, metal-rich disks that end up exciting higher orbital eccentricities in the planets \citep{Dawson:2013}.

More recent studies have also shed light on the links between metallicity, planet radius, and eccentricity. For example, \cite{Gilbert:2025} measured orbital eccentricities for a sample of $\sim$1640 {\it Kepler} Sun-like systems, and they found that planets approximately larger than Neptune ($R_p \gtrsim 3.5R_{\oplus}$) exhibit elevated eccentricities. Because larger planets also typically require higher stellar metallicities to form, a positive correlation is also expected between eccentricity and [Fe/H], which \cite{Gilbert:2025} establish with significant confidence (3$\sigma$) for sub-Saturns, and more modestly for Jovians, where the trend is consistent with a flat line within 1$\sigma$. Expanding this analysis to the M-dwarf planetary population, \cite{Sagear:2026} also found evidence for a transition towards higher eccentricities for planets larger than 3.5$R_{\oplus}$.

\begin{figure}[ht!]
\vspace{.1in}
\centering
\includegraphics[width=1\linewidth]{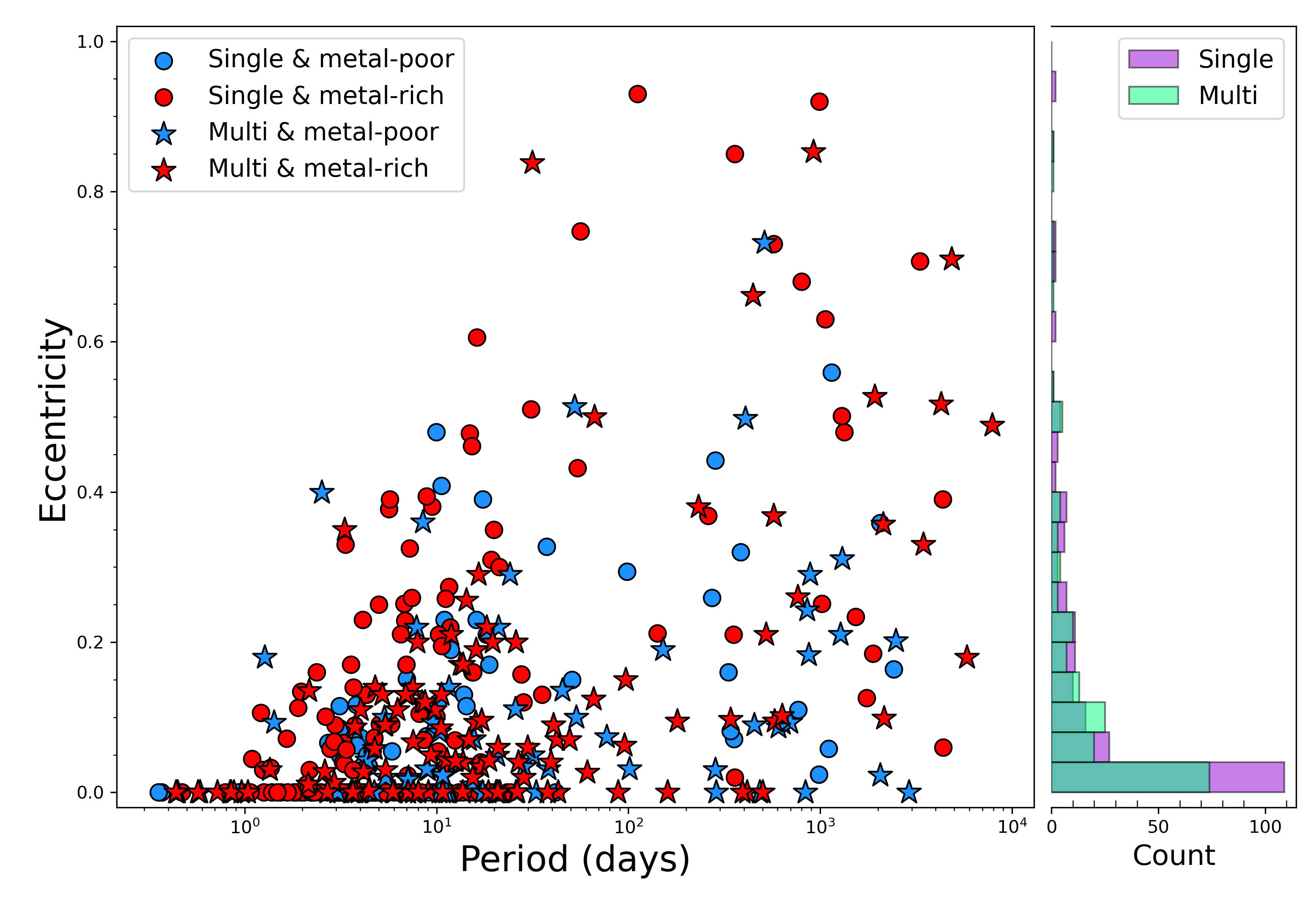}
%\vspace{-.25in}
\caption{Period-eccentricity distribution of the stars in our sample with RV-measured mass planets. The circles represent stars with a single known planet and the star markers are stars with multiple planets. Red points represents stars with [Fe/H] $>$ 0 and blue ones are those with [Fe/H] $\leq$ 0. There is an apparent lack of metal-poor ([Fe/H] $\leq$ 0) stars at high orbital eccentricities ($e>$ 0.5), as previously shown by \cite{Dawson:2013, Winn:2015}. The side histogram shows the distributions of singles and multis in purple and green, respectively.}
\label{fig:period_eccentricity} 
\end{figure}

Here, we extend that analysis using our homogeneously derived [Fe/H] metallicities and show an updated version of the plot using only planets with RV masses (which have the most reliable eccentricity measurements) in Figure~\ref{fig:period_eccentricity}.  The points are color-coded by metallicity and the symbols denote planet multiplicity. As in the original RV sample from \cite{Winn:2015}, a clear deficit of metal-poor stars (i.e., a lack of blue points) is visible at higher eccentricities, reinforcing the trends observed by \cite{Dawson:2013} and \cite{Winn:2015} in a much larger and homogeneous sample.

Performing the same analysis in terms of [$\alpha$/Fe] does not reveal similarly insightful trends, in part because the split between $\alpha$-poor and $\alpha$-enriched is less well-defined than it is for [Fe/H], and the sample of stars with high [$\alpha$/Fe] values is small. However, we note that the planets orbiting $\alpha$-enriched ($\geq$0.2 dex) hosts in our sample (24 planets around 13 stars) have low eccentricities ($e < 0.3$), long periods (with a median of $P = 57$ days), and large radii (median of $R = 8 R_{\oplus}$).

\subsubsection{Stellar Rotation and Planet Multiplicity} \label{subsec:metallicity_rotation}

We investigated the correlation between stellar projected rotational velocities ($v_{*}$sin$i$) and planet multiplicity. Previous studies have reported mixed results. \citet{Weiss:2018} analyzed Sun-like stars in the California-Kepler Survey and reported no statistically significant differences in the $v_{*}$sin$i$ distributions of singles and multis. \cite{Ballard:2016} found modest evidence ($\approx$2$\sigma$) that multi-planet systems around M dwarfs orbit slightly faster rotators than single-planet systems.
In our sample of FGK stars, we observe a weak, tentative trend in the opposite sense: single-planet stars have modestly higher $v_{*}$sin$i$ (median of 4 km/s) than multi-planet systems (median of 3.8 km/s). Applying the KS and AD two-sample tests to stars cooler than the Kraft break ($T_{\rm eff}$ $\leq$ 6200 K) yields $p$-values of 0.0015 (AD) and 0.24 (KS), indicating a potentially  statistically significant difference. However, these results must be interpreted with caution. First, the rotation velocities measured in this work are projected velocities ($v_{*}$sin$i$), not the true equatorial speeds ($v_{true}$). Because $v_{true}$ depends on the unknown inclination angle $i$ of each stellar rotation axis, projection effects introduce significant scatter. In addition, we note that half of a TRES resolution element corresponds to 3.4~km/s; i.e., rotational broadening signals that are smaller than that are not trustworthy given the instrumental broadening.
Additional sources of bias further complicate the interpretation. Rapidly rotating stars exhibit broadened spectral lines, making planet detection, especially of multiple planets, more difficult; this selection effect would naturally produce a dearth of high-$v_{*}$sin$i$ multi-planet hosts. Uncertainties in planet multiplicity also play a role, since some apparently single systems may in fact host additional undetected planets.
A definitive assessment of whether true stellar rotation correlates with planetary system architecture would require addressing these detection and completeness biases, as well as obtaining independent constraints on stellar inclinations for a large ensemble of systems.

\begin{deluxetable*}{lcccccccccc}[ht]
\tablenum{1} \tablecolumns{6}
\tablecaption{Stellar Properties\label{tab:stellarprops}}
\tablehead{
\colhead{TIC ID} & \colhead{$T_{\rm eff}$ (K)} & \colhead{$\log g_*$} & \colhead{[Fe/H]} & \colhead{[$\alpha$/Fe]} & \colhead{Radius ($R_{\odot}$)} & Luminosity ($L_{\odot}$)& \colhead{$v_{\star}$sin$i$} &  \colhead{$\rm U_{LSR}$} & \colhead{$\rm V_{LSR}$} & \colhead{$\rm W_{LSR}$}}
\startdata
1003831 & 5640 $\pm$ 101 & 4.39 $\pm$ 0.09 &0.19 $\pm$ 0.04 & 0.04 $\pm$ 0.03 &  1.06 $\pm$ 0.13 & 1.02 $\pm$ 0.18 & 4.3 $\pm$ 3.4 &  31.22	& $-$1.74 & $-$1.78 \\
229846907 & 4836 $\pm$ 100 & 2.50 $\pm$ 0.14 & $-$0.32 $\pm$ 0.10 & 0.22 $\pm$ 0.15 & 13.35  $\pm$ 3.60 & 87.47	$\pm$ 51.61 & 4.6 $\pm$ 3.4& $-$22.40	& 22.24	&0.77 \\
53458803 & 5803 $\pm$ 101 & 4.36 $\pm$ 0.09 & 0.39 $\pm$ 0.04 & 0.03 $\pm$ 0.03 & 1.15 $\pm$	0.14 & 1.35	$\pm$ 0.24 & 4.6 $\pm$	3.4 & 12.94 &	15.54 &	$-$25.47\\
\enddata
\tablenotetext{a}{These values are our results in the {\tt MINESweeper} (MS) mode; i.e., with isochrones. This table is available in its entirety in machine-readable format. We additionally provide RA and Dec and TD/D ratios for every target.}
\end{deluxetable*}

\section{Conclusions} \label{conclusions}

In this paper, we uniformly characterized a sample of 625 hosts to 859 confirmed exoplanet systems with declinations above --40$^{\circ}$ using archival high-resolution optical spectra from the TRES spectrograph. We modeled our data with the spectral neural-net code {\tt uberMS} in two modes: one that includes spectroscopy, photometry and stellar atmospheres, and another that additionally includes stellar isochrones to homogenously estimate precise and accurate stellar effective temperatures, surface gravities, iron and $\alpha$-process enrichment, stellar radii, luminosities, and projected rotational velocities.

Based on our [Fe/H] and [$\alpha$/Fe] values, we identified 51 planet hosts that could be members of the galactic thick disk based on their chemistry, and 14 stars that have a high likelihood of kinematic thick-disk membership, with 7 targets consistent with the thick disk based on both chemistry and kinematics. We also note that there are 5 additional targets that have reasonably high likelihoods of belonging to the thick disk, with $1< P(TD/D) <10$, and which are chemically consistent with the thick disk, suggesting an even greater overlap of 12. Further analysis and follow-up observations into these potential thick disk members could shed light into this rare population of planet hosts and into the diversity of planet-forming environments in our galaxy.

We found that the metal-poor ([Fe/H] $\leq$ 0)  hosts to giant planets have significantly more elevated [$\alpha$/Fe] than those with super-solar metallicities, with AD and KS $p$-values of $p = 1\times10^{-5}$ and $p = 6\times10^{-8}$, respectively. This suggests that $\alpha$-enhancement may partially help compensate for low-Fe and thus [$\alpha$/Fe] may play an important role in giant planet formation. We also explored the differences between the [$\alpha$/Fe] distributions of stars with a single known planet versus those with multiple, and we found modest evidence that $\alpha$-rich hosts are more likely to host multiple planets, with AD and KS $p$-values of 0.002 and 0.19, respectively.

This catalog is a useful resource for stellar and planetary demographics and lays the groundwork for future statistical studies to probe other trends between host star chemistry and planetary properties.
 
\section*{Acknowledgments}
We thank the anonymous referee for helpful comments that improved this manuscript. We are grateful to Allyson Bieryla, Samuel Yee, Samuel Quinn, Anusha Pai Asnodkar, and Caprice Phillips for insightful conversations. R.R.M. is supported by a Harvard Postdoctoral Fellowship for Future Faculty Leaders. V.D. acknowledges support from the National Science Foundation Graduate Research Fellowship under Grant No.\ DGE1745303. E.P. is supported by a Juan Carlos Torres Postdoctoral Fellowship at the Massachusetts Institute of Technology. The authors wish to thank TRES team members Perry Berlind, Michael Calkins, Gilbert Esquerdo, Pascal Fortin, Jessica Mink, and Andrew Szentgyorgyi, for the development and operation of the TRES instrument. We also wish to thank Guillermo Torres, Benjamin Montet, Jason Curtis, Katja Poppenhaeger, and Jennifer Winters, for part of the TRES observations used in this project.
This paper includes data collected by the TESS mission. Funding for the TESS mission is provided by the NASA's Science Mission Directorate. This research has also made use of the Exoplanet Follow-up Observation Program website, which is operated by the California Institute of Technology, under contract with the National Aeronautics and Space Administration under the Exoplanet Exploration Program.
This work has made use of data from the European Space Agency (ESA) mission
{\it Gaia} (\url{https://www.cosmos.esa.int/gaia}), processed by the {\it Gaia}
Data Processing and Analysis Consortium (DPAC, \url{https://www.cosmos.esa.int/web/gaia/dpac/consortium}). 

\vspace{5mm}

\facilities{FLWO:1.5m (TRES), {\it Gaia}.}

\software{\texttt{uberMS} (\citealt{Ting:2019, Cargile:2020}, Cargile et al. in prep.),   \texttt{Astropy} \citep{Astropy2022}, \texttt{JAX} \citep{Bradbury2018}, \texttt{Matplotlib} \citep{Hunter2007}, \texttt{NumPy} \citep{Harris2020}, \texttt{NumPyro} \citep{Phan2019}, \texttt{pandas} \citep{Reback2021}, \texttt{SciPy} \citep{Scipy2020}.}

\bibliography{aastex7}{}
\bibliographystyle{aasjournalv7}

\clearpage

\appendix
%\twocolappendix
\section{Long Tables}
\restartappendixnumbering
\begin{deluxetable}{lllcc}[h]
\tabletypesize{\footnotesize}
\tablecolumns{7}
\tablewidth{0pt}
 \tablecaption{Fitting parameters and priors used with \texttt{uberMS} \label{tab:priors}}
 \tablehead{
 \colhead{Parameter} &
 \colhead{Variable} &
 \colhead{Unit} &
 \colhead{Initial Value} &
 \colhead{Prior}}
\startdata
\multicolumn{5}{c}{\emph{Spectroscopic parameters}} \\
Resolution (per order) & lsf\_\{ii\} & --- & 44000 & $\mathcal{N}(44000, 1000)$; truncated at 40000 and 53000\\
Spectroscopic jitter (per order) & specjitter\_\{ii\} & --- & 0.015 & fixed \\
Radial velocity (per order) & vrad\_\{ii\} & kms$^{-1}$ & 0.0 & $\mathcal{U}(-5, 5)$\\
Normalization (per order) & pc0\_\{ii\} & --- & 1.0 & $\mathcal{N}(1.0, 0.01)$; truncated at 0.75 and 2.0 \\
Linear trend (per order) & pc1\_\{ii\} & --- & 0.0 & $\mathcal{U}(-0.2, 0.2)$ \\
Stellar broadening & vstar & kms$^{-1}$ & 2.0 & $\mathcal{U}(0, 250)$\\
Microturbulence & vmic & kms$^{-1}$ & \cite{Bruntt:2012} & \cite{Bruntt:2012} empirical relation \\
\\
\multicolumn{5}{c}{\emph{Photometric parameters}} \\
Extinction & Av & mag & $A_{\rm v,calc}$/2 & $\mathcal{U}{(0, A_{\rm v,calc}}$) using \cite{Vergely:2022} \\
Distance & dist & pc & {\it Gaia} DR3 & $\mathcal{U}(-5\sigma, 5\sigma)$ using parallax errors from {\it Gaia} DR3 \\
Photometric jitter & photjitter & mag & 0.03 & $\mathcal{N}(0.02, 0.01)$; truncated at 0. and 0.1\\
\\
\multicolumn{5}{c}{\emph{TP-mode parameters}} \\
Effective temperature & Teff & K & 5000 & $\mathcal{U}(2500, 12000)$\\
Surface gravity & log(g) & dex & 4.5  & $\mathcal{U}(0.5, 5.5)$\\
Iron abundance & [Fe/H] & dex & 0.0  & $\mathcal{U}(-4.0, 0.5)$ \\
Alpha enrichment & [$\alpha$/Fe] & dex & 0.0  & $\mathcal{U}(-0.2, 0.6)$ \\
Stellar radius (log10) & log(R) & dex; $R$ in R$_\odot$ &  0.0 & $\mathcal{U}(-3, 3)$ \\
\\
\multicolumn{5}{c}{\emph{MS-mode parameters}} \\
Equivalent evolutionary phase & EEP & --- & 250 & $\mathcal{N}(250, 50)$; truncated at 200 and 600\\
Stellar age (log10) & log(Age) & dex; Age in years & Latent variable & dsigmoid(80., 8., $-40$, 10.15, 6.0, 11.0)\\
Initial iron abundance & initial\_[Fe/H] & dex & 0.0 & $\mathcal{U}(-4, 0.5)$\\
Initial alpha enrichment & initial\_[$\alpha$/Fe] & dex & 0.0 & $\mathcal{U}(-0.2, 0.6)$\\
Initial stellar mass & initial\_Mass & M$_\odot$ & 1.0 & $\mathcal{U}(0.5, 3.0)$
\enddata
\centering{
}
\end{deluxetable}

\startlongtable
\begin{deluxetable}{lccccccc}
\tablenum{2}
\tablecolumns{8}
\tablecaption{Potential Thick Disk Stars\label{tab:thickdisk}}
\tablehead{
\colhead{TIC ID} &
\colhead{$T_{\rm eff}$ (K)} &
\colhead{$\log g_*$} &
\colhead{[Fe/H]} &
\colhead{[$\alpha$/Fe]} &
\colhead{TD/D ratio} &
\colhead{Disk Classification Method} &
\colhead{Sigmas above thick-disk line} 
}
\startdata
445811628 & 4904 & 4.48 & -0.63 & 0.36  & 3895558.65 & chemical \& kinematic & 5.3  \\
14227229  & 5823 & 4.25 & -0.44 & 0.30   & 17793.89   & chemical \& kinematic & 3.6  \\
432247186 & 5324 & 4.52 & -0.60  & 0.31  & 536.76     & chemical \& kinematic & 3.4  \\
164456037 & 5927 & 4.01 & -0.37 & 0.26  & 486.03     & chemical \& kinematic & 2.8  \\
138297129 & 5653 & 4.40  & -0.15 & 0.18  & 92.26      & chemical \& kinematic & 1.3  \\
366411016 & 5174 & 4.55 & -0.62 & 0.20   & 59.50       & chemical \& kinematic & 0.0 \\
20484888  & 5770 & 4.15 & -0.44 & 0.28  & 11.05      & chemical \& kinematic & 2.3  \\
124573851 & 5628 & 4.44 & -0.03 & 0.04  & 70.32      & kinematic             & -2.6 \\
232540264 & 5720 & 4.38 & -0.11 & 0.13  & 41.05      & kinematic             & -0.1 \\
219015370 & 5753 & 4.17 & 0.31  & 0.07  & 26.02      & kinematic             & -1.6 \\
303432813 & 4832 & 4.49 & 0.16  & -0.01 & 24.53      & kinematic             & -4.2 \\
342642208 & 5618 & 4.20  & 0.36  & 0.02  & 18.72      & kinematic             & -3.3 \\
122515955 & 5269 & 4.19 & -0.01 & 0.10   & 15.51      & kinematic             & -0.7 \\
356473034 & 5240 & 4.39 & 0.38  & 0.03  & 12.20       & kinematic             & -3.0    \\
333657795 & 5616 & 4.33 & -0.03 & 0.25  & 8.59       & chemical              & 4.1  \\
377780790 & 5719 & 4.37 & -0.08 & 0.14  & 8.25       & chemical              & 0.3  \\
25958378  & 5182 & 4.50  & -0.38 & 0.34  & 5.35       & chemical              & 5.3  \\
377064495 & 5326 & 4.47 & -0.31 & 0.24  & 1.29       & chemical              & 2.6  \\
709015    & 5496 & 4.48 & -0.41 & 0.25  & 1.04       & chemical              & 2.2  \\
349445839 & 4829 & 3.23 & -0.16 & 0.21  & 0.87       & chemical              & 0.5  \\
164557694 & 5789 & 4.09 & -0.38 & 0.24  & 0.87       & chemical              & 1.8  \\
121327187 & 5788 & 4.12 & -0.35 & 0.24  & 0.85       & chemical              & 2.1  \\
121124617 & 5456 & 4.51 & -0.41 & 0.23  & 0.79       & chemical              & 1.8  \\
158215347 & 5733 & 4.31 & -0.01 & 0.13  & 0.65       & chemical              & 0.3  \\
399402994 & 5945 & 4.29 & -0.29 & 0.20   & 0.48       & chemical              & 1.2  \\
392469577 & 5695 & 4.46 & -0.17 & 0.17  & 0.25       & chemical              & 0.9  \\
209459275 & 5936 & 4.28 & -0.24 & 0.21  & 0.20        & chemical              & 1.8  \\
268532343 & 5617 & 3.82 & -0.18 & 0.16  & 0.19       & chemical              & 0.1  \\
138969239 & 5840 & 4.04 & -0.27 & 0.18  & 0.19       & chemical              & 0.6  \\
408316773 & 4971 & 3.15 & -0.29 & 0.20   & 0.19       & chemical              & 0.3  \\
269701147 & 5394 & 4.42 & -0.13 & 0.22  & 0.15       & chemical              & 2.8  \\
158839841 & 5795 & 4.33 & -0.17 & 0.16  & 0.13       & chemical              & 0.5  \\
275573429 & 5778 & 4.29 & -0.22 & 0.32  & 0.11       & chemical              & 4.8  \\
147677253 & 5866 & 4.24 & -0.41 & 0.26  & 0.09       & chemical              & 2.7  \\
120420633 & 5311 & 3.81 & -0.15 & 0.15  & 0.07       & chemical              & 0.1  \\
281731203 & 5522 & 4.06 & 0.38  & 0.15  & 0.06       & chemical              & 1.0     \\
57983992  & 4888 & 3.39 & 0.05  & 0.14  & 0.05       & chemical              & 0.1  \\
28763463  & 4809 & 2.48 & -0.30  & 0.22  & 0.05       & chemical              & 0.4  \\
1713457   & 4918 & 3.15 & -0.15 & 0.17  & 0.04       & chemical              & 0.2   \\
115010361 & 5581 & 4.44 & -0.16 & 0.23  & 0.04       & chemical              & 2.7   \\
158114249 & 5604 & 3.97 & -0.36 & 0.25  & 0.04       & chemical              & 0.6  \\
103751498 & 5732 & 4.22 & 0.06  & 0.13  & 0.04       & chemical              & 0.3  \\
49430557  & 4810 & 2.52 & -0.26 & 0.23  & 0.04       & chemical              & 0.5  \\
229846907 & 4836 & 2.51 & -0.32 & 0.22  & 0.03       & chemical              & 0.4  \\
399334674 & 5057 & 3.23 & -0.30  & 0.20   & 0.03       & chemical              & 0.3  \\
178941218 & 5775 & 4.49 & -0.08 & 0.17  & 0.03       & chemical              & 1.2  \\
26814296  & 6262 & 4.39 & -0.07 & 0.16  & 0.03       & chemical              & 0.8  \\
113329742 & 4937 & 3.10  & -0.21 & 0.18  & 0.03       & chemical              & 0.2  \\
456826468 & 4895 & 3.21 & -0.07 & 0.15  & 0.03       & chemical              & 0.1  \\
354489950 & 4950 & 2.91 & -0.13 & 0.20   & 0.03       & chemical              & 0.4  \\
288329878 & 5084 & 2.86 & -0.13 & 0.20   & 0.03       & chemical              & 0.4  \\
38948719  & 4944 & 3.35 & -0.08 & 0.14  & 0.02       & chemical              & 0.1  \\
413853880 & 4931 & 3.36 & -0.04 & 0.14  & 0.02       & chemical              & 0.1   \\
9815387   & 4829 & 2.56 & -0.04 & 0.23  & 0.02       & chemical              & 0.7   \\
344627954 & 4815 & 3.21 & -0.04 & 0.13  & 0.02       & chemical              & 0.0  \\
251095345 & 4805 & 2.59 & -0.13 & 0.22  & 0.02       & chemical              & 0.6  \\
20096620  & 5673 & 4.02 & 0.38  & 0.13  & 0.02       & chemical              & 0.3  \\
53873088  & 4825 & 3.13 & -0.03 & 0.14  & 0.02       & chemical              & 0.1 \\
\enddata
\tablenotetext{a}{These values are our results in the MS mode. TD/D ratios were computed following \citet{Bensby:2003}. TD/D $\geq$ 10 indicate high thick$-$disk membership likelihood. Stars are sorted by decreasing TD/D ratios.}
\end{deluxetable}

\end{document}